\documentclass[12pt]{amsart}
\usepackage{amsmath}
\usepackage{amssymb}
\usepackage[latin1]{inputenc}
\usepackage[T1]{fontenc}
\newtheorem{lemma}{Lemma}
\newtheorem{theorem}{Theorem}
\newtheorem{proposition}{Proposition}
\newtheorem{corollary}{Corollary}

\newcommand{\tr}{{\rm Tr }}

\newcommand{\hog}[1]{\beta(#1)}

\newcommand{\aut}{{\rm Aut }}

\newcommand{\bra}{\langle}
\newcommand{\ket}{\rangle}
\newcommand{\vp}{\varphi}

\newcommand{\C}{\mathbb{C}}
\newcommand{\N}{\mathbb{N}}

\newcommand{\R}{\mathbb{R}}
\newcommand{\T}{\mathbb{T}}
\newcommand{\be}{\begin{equation}}
\newcommand{\eeq}{\end{equation}}
\newcommand{\bet}{\begin{equation*}}
\newcommand{\eeqt}{\end{equation*}}
\newcommand{\bea}{\begin{eqnarray}}
\newcommand{\eeqa}{\end{eqnarray}}
\newcommand{\beat}{\begin{eqnarray*}}
\newcommand{\eeqat}{\end{eqnarray*}}
\newcommand{\goesto}{\longrightarrow}

\newcommand{\hil}{\mathcal{H}}

\newcommand{\hU}{\mathcal{U}}
\newcommand{\hT}{\mathcal{T}}

\newcommand{\hHS}{\mathcal{HS}}

\newcommand{\hB}{\mathcal{B}}

\newcommand{\hF}{\mathcal{F}}
\newcommand{\hK}{\mathcal{K}}
\newcommand{\lambdat}{\tilde{\lambda}}

\begin{document}
\title{Covariant observables on a nonunimodular group}
\author{J. Kiukas}
\address{Jukka Kiukas,
Department of Physics, University of Turku,
FIN-20014 Turku, Finland}
\email{jukka.kiukas@utu.fi}
\begin{abstract}
It is shown that the characterization of covariant positive operator measures on nonunimodular locally compact groups can be obtained by
using vector measure theoretic methods, without an application of Mackey's imprimitivity theorem.
\end{abstract}
\maketitle

\section{Introduction}
Covariant positive operator measures have an important physical significance, as they represent
phase space translation covariant quantum mechanical observables (see e.g. \cite{HolevoII}).
On the other hand, they can be used in quantization. Namely, in the context of a locally compact, second countable topological group $G$ with
a Haar measure $\lambda$, each positive normal covariant map $\Gamma:L^\infty(G,\lambda)\to L(\hil)$, with $L(\hil)$ the set of bounded
operators on some separable Hilbert space $\hil$, is eligible to represent a quantization procedure \cite{Werner,ncqm}.
The maps of this kind correspond to covariant positive operator measures
via the association $\hB(G)\ni B\mapsto \Gamma(\chi_B)\in L(\hil)$, where $\hB(G)$ is the Borel $\sigma$-algebra of $G$ and $\chi_B$ the
characteristic function of the set $B$.

Since covariant observables are essential in quantum
mechanics, they have been studied quite extensively. The canonical examples of covariant observables are constructed e.g. in \cite{Davies}, and
there are (at least) two completely different ways to obtain their characterization: a group theoretical approach \cite{Cassinelli}, and
a direct approach \cite{Holevo,Werner,ncqm} based on the theory of integration with respect to vector measures. The latter approach was presented by Werner in
\cite{Werner} in the context where $G=\R^{2n}$, and it was generalized to the case of a unimodular group in \cite{ncqm}. The assumption of unimodularity
was quite essential in the proof, and there is no trivial way of getting rid of it. However, the group theoretical
approach of \cite{Cassinelli}, which is based on a generalization of Mackey's imprimitivity theorem, works also in the nonunimodular case,
which raises a question of whether there is something "essentially group theoretical" in the characterization.

The purpose of this paper is to show that with some modifications, the direct approach indeed works also in the nonunimodular case.
As in \cite{Werner} and \cite{ncqm}, the proof relies on the fact that the Banach space of trace class operators on a separable Hilbert space has the Radon-Nikod\'ym
property. The essential difference is that Lemma 3.1 of \cite{Werner} and its generalization \cite[Lemma 2]{ncqm} no longer hold if $G$ is
not unimodular. That result must be replaced by a weaker version,
which is a consequence of the classical work of Duflo and Moore \cite{Duflo} concerning square integrable representations.

In \cite{Werner} and \cite{ncqm}, the characterization of the positive normal covariant maps was obtained first, and the characterization
for covariant observables was then deduced
from it. It this paper, however, we find it convenient to use observables from the beginning.

\section{Preliminaries}

Let $\hil$ be a separable Hilbert space with the inner product $\bra \cdot|\cdot\ket$ linear in the \emph{second} argument.
We let $L(\hil)$, $\hHS(\hil)$, and $\hT(\hil)$ denote the Banach spaces of bounded, Hilbert-Schmidt, and
trace class operators on $\hil$, respectively. We use the symbols $\|\cdot\|$, $\|\cdot\|_{HS}$ and $\|\cdot\|_{\tr}$ for the norms of these
spaces. The symbol $\hU(\hil)$ stands for the set of unitary operators on $\hil$, and
$\T$ denotes the group of complex numbers with modulus one.

For a linear, not necessarily bounded operator $A$ on $\hil$, the symbol $D(A)$ denotes the domain
of definition of $A$. If $\vp\in \hil$, we use the symbol $|\vp\ket\bra \vp|$ for the operator
$\psi\mapsto \bra \vp|\psi\ket \vp$.

Let $\aut (\hT(\hil))$ denote the group of positive, trace-norm preserving linear
bijections from $\hT(\hil)$ onto itself. The group $\aut (\hT(\hil))$ is equipped with the weak topology
given by the set of functionals $u\mapsto \tr[A u(T)]$, where $A\in L(\hil)$, $T\in \hT(\hil)$.
For $u\in \aut(\hT(\hil))$, the adjoint map $u^*:L(\hil)\to L(\hil)$ restricted to $\hT(\hil)$ is equal to $u^{-1}$.
It follows from the Wigner theorem that for each $u\in \aut(\hT(\hil))$ there is either
a unitary or an antiunitary operator $U$, such that $u(T) = UTU^*$ for all $T\in\hT(\hil)$. Clearly then $u^{-1}$
is given by $u^{-1}(T) = U^*TU$.

Let $G$ be a locally compact (not necessarily unimodular) second countable (Hausdorff) topological group, with a
left Haar measure $\lambda$. Let $\lambdat$ denote the right Haar measure $B\mapsto \lambda(B^{-1})$, and
let $\Delta$ be the modular function so that $\lambdat(B) = \int_B \Delta(g)^{-1} d\lambda(g)$ for all Borel sets $B$.
Now $\lambda$ and $\lambdat$ have the same null sets, and both of them are $\sigma$-finite.
We let $\hB(G)$ denote the Borel $\sigma$-algebra of the subsets of $G$.

The following definition was used in \cite{ncqm} for the measurability of a vector valued function. It is sometimes called
strong measurability (see e.g. \cite[p. 41]{Diestel}).

A function $f$ defined on $G$ and having values in
some Banach space is said to be \emph{$\lambda$-measurable}, if for each $B\in \hB(G)$ of finite $\lambda$-measure there is a sequence of
$\lambda$-simple functions converging to $\chi_Bf$ in $\lambda$-measure (or, equivalently, there is a sequence
of $\lambda$-simple functions which converges $\lambda$-almost everywhere to $\chi_Bf$) \cite[pp. 106, 150]{Dunford}.
In the case where the value space of $f$ is separable (in particular, if $f$ is scalar-valued), $\lambda$-measurability is
equivalent to the measurability with respect to the Lebesgue extension of the $\sigma$-algebra $\hB(G)$ with respect
to $\lambda$ \cite[p. 148]{Dunford}. Measurability with respect to $\lambdat$ is of course defined in the same way.
Since $G$ is $\sigma$-compact, with $\lambda$ and $\lambdat$ having the same null sets and being finite on compact sets,
it follows that a vector valued function is $\lambda$-measurable if and only if it is $\lambdat$-measurable.

We let $L^\infty(\Omega, \lambda)$ denote the Banach space of (equivalence classes of) complex valued,
$\lambda$-measurable, $\lambda$-essentially bounded functions.

An \emph{observable} is a positive normalized
operator measure, i.e. a positive operator valued map $E:\hB(G)\to L(\hil)$, which is weakly (or, equivalently, strongly) $\sigma$-additive,
and such that $E(G) = I$.

In \cite{ncqm}, the starting point was to introduce a continuous homomorphism $\beta:G\to\aut(\hil)$, with the property
that $\int \tr[P_1\beta(g)(P_2)] d\lambda(g) = d$ for all one dimensional projections $P_1$ and $P_2$, with $0<d<\infty$ a fixed constant.
If $G$ is connected, and not unimodular, there are no such homomorphisms (see Lemma \ref{trlemma} and the Remark (c) following Lemma \ref{easylemma}).
Instead, we have to start with the concept of projective unitary representation, which is used extensively in quantum mechanics (see e.g. \cite[Chapter VII]{Varadarajan}).
It is defined as follows.

\

A map $U:G\to \hU(\hil)$ is a \emph{projective unitary representation}, if
\begin{itemize}
\item[(i)] the map $g\mapsto U(g)$ is weakly Borel, i.e. $g\mapsto \bra \psi |U(g)\vp\ket$ is a Borel function for all $\psi, \vp\in \hil$;
\item[(ii)] $U(e)=I$, where $e$ is the neutral element of $G$;
\item[(iii)] there is a Borel map $m:G\times G\to \T$, such that $U(gh) = m(g,h)U(g)U(h)$ for all $g,h\in G$.
The map $m$ (clearly unique) is called the \emph{multiplier} of $U$.
\end{itemize}

\

The irreducibility of a projective unitary representation is defined in the same way as in the case of ordinary unitary representations.
For each projective unitary representation $U:G\to\hU(\hil)$, we let $\beta_U: G\to \aut(\hT(\hil))$ be the map given by
$\beta_U(g)(S) = U(g) SU(g)^*$.

\

We need the following simple result, which is, of course, well known.
\begin{lemma}\label{easylemma}
Let $U:G\to \hU(\hil)$ be a projective unitary representation. Then the map $\beta_U$ is a group homomorphism, with the property
that for each $A\in L(\hil)$ and $S\in \hT(\hil)$, the map $g\mapsto \tr[A\beta_U(g)(S)]$ is a Borel function.
\end{lemma}
\begin{proof} It is obvious that $\beta_U$ is a group homomorphism.  
Since the map $g\mapsto U(g)$ is weakly Borel, all the maps $g\mapsto \bra \psi |U(g) \vp\ket\bra U(g)\vp'|\psi'\ket$, with $\vp,\vp',\psi,\psi'\in\hil$, are Borel functions.
The separability of $\hil$ implies that the function $g\mapsto \tr[A\beta_U(g)(S)]$ is a (pointwise) limit of linear combinations of such maps, and
hence it is Borel.
\end{proof}
\noindent {\bf Remark.} The following remarks provide some additional, basically well-known facts on the map $\beta_U$ associated
with a projective unitary representation $U$.
However, we do not need to use these facts in this paper; they serve as a connection to the paper \cite{ncqm}, where a continuous
homomorphism $\beta:G\mapsto \aut(\hT(\hil))$ had a central role.
\begin{itemize}
\item[(a)] Since $\hil$ is separable, containing a countable dense set $M$, it follows that the whole topology of $\aut(\hT(\hil))$ is given by the countable
family of functionals
\bet
\hF = \{u\mapsto \tr[|\psi\ket\bra\psi|u(|\vp\ket\bra \vp|)]\mid \psi, \vp\in M\}.
\eeqt
This is a consequence of the fact that while the Wigner isomorphism $\Sigma \ni [U]\mapsto \beta_U\in \aut(\hT(\hil))$
(see e.g. \cite[Chapter 2]{Symmetry}) is continuous when $\aut(\hT(\hil))$ is equipped with the usual topology,
its inverse is continuous even when $\aut(\hT(\hil))$ is considered with the (a priori weaker) topology given by the family $\hF$; see
the proof of \cite[Proposition 10]{Symmetry}. Here $\Sigma$ denotes the equivalence classes of all unitary and antiunitary operators on $\hil$,
with the equivalence relation being equality up to a phase factor. It follows that $\aut(\hT(\hil))$ is second countable, and the
$\beta_U$ of the preceding lemma is a Borel function (with respect to the Borel structure given by the associated topologies).
\item[(b)] A classical theorem of von Neumann states that a Borel homomorphism from a second countable
topological group to another is continuous, provided that the former group is locally compact (see e.g. \cite[p. 181]{Varadarajan}).
Hence, the $\beta_U$ of the preceding lemma
is in fact continuous (so that each map $g\mapsto \tr[A\beta_U(g)(T)]$, with $A\in L(\hil)$ and $T\in \hT(\hil)$, is such).
\item[(c)] As mentioned before, each $u\in \aut(\hT(\hil))$ is of the form $u(T)=UTU^*$ for some unitary or antiunitary operator $U$ on $\hil$.
It follows that, in the case where $G$ is connected, each continuous homomorphism $\beta:G\to \aut(\hT(\hil))$ is of the form $\beta_U$ for some
projective unitary representation $U:G\to \hU(\hil)$. Moreover, it is easily seen that the irreducibility of $U$ is equivalent to the following condition:
\be\label{irrcond}
\text{for all one-dimensional projections } P_1, P_2 \text{, there is } g\in G, \text{ with } \tr[P_1\beta_U(g)(P_2)]\neq 0.
\eeq
Thus, in the case where $G$ is connected, the irreducible projective representations $U:G\to \hU(\hil)$
are in a one-to-one correspondence with the continuous homomorphisms $\beta: G\to \aut(\hT(\hil))$ with the property \eqref{irrcond}.
\end{itemize}

\

The following result can be extracted from some of the proofs in \cite{ncqm}. We give the proof here for clarity. 

\begin{lemma} \label{measurabilitylemma} Let $U:G\to \hU(\hil)$ be a projective unitary representation, and
$v:G\to \hT(\hil)$ a $\lambda$-measurable function. Then also the functions $g\mapsto \beta_U(g)(v(g))$ and
$g\mapsto \beta_U(g^{-1})(v(g))$ are $\lambda$-measurable.
\end{lemma}
\begin{proof}
First we notice that for each $S\in \hT(\hil)$, the map $g\mapsto \beta_U(g^{-1})(S)$ is $\lambda$-measurable.
Indeed, let $S\in \hT(\hil)$. It follows from Lemma \ref{easylemma} that
$G\ni g\mapsto w^*(\beta_U(g^{-1})(S))\in \C$ is a Borel function, and hence $\lambda$-measurable for each $w^*\in \hT(\hil)^* \cong L(\hil)$.
Since $\hT(\hil)$ is separable (see e.g. \cite[Lemma 5]{ncqm}), this implies by \cite[p. 149]{Dunford} that the map
$g\mapsto \beta_U(g^{-1})(S)$ is $\lambda$-measurable.

Now let $B\in\hB(G)$ be such that $\lambda(B) <\infty$. Since $v$ is $\lambda$-measurable, there is a sequence
$(\tilde{v}_n)$ of $\lambda$-simple functions vanishing outside $B$ and converging $\lambda$-a.e. to $\chi_Bv$.
Since the map $g\mapsto \beta_U(g^{-1})(S)$ is $\lambda$-measurable for each $S\in \hT(\hil)$, also
the functions $g\mapsto \beta_U(g^{-1})(\tilde{v}_n(g))$, are $\lambda$-measurable.
Now $\beta_U(g^{-1})(\tilde{v}_n(g))\goesto \chi_B(g) \beta_U(g^{-1})(v(g))$ for $\lambda$-almost all $g$, because
each $\beta_U(g^{-1})\in \aut(\hil)$ is continuous, so the limit is $\lambda$-measurable \cite[p. 150]{Dunford}.
Thus also $g\mapsto \beta_U(g^{-1})(v(g))$ is $\lambda$-measurable.

The $\lambda$-measurability of $g\mapsto \beta_U(g)(v(g))$ is established similarly.
\end{proof}

A projective representation $U:G\to \hU(\hil)$ is called \emph{square integrable}, if there exist nonzero vectors $\vp,\psi\in\hil$, such
that the function $g\mapsto |\bra \psi |U(g)\vp\ket|^2$ is $\lambda$-integrable. The theory of square integrable representations is usually
presented only in the context of ordinary unitary representations. An essential
result is that for an irreducible, square integrable representation $U$, there exists a unique, densely defined, injective, positive selfadjoint operator
$K$, called the \emph{formal degree} of $U$, such that $U(g) K = \Delta(g)^{-1}KU(g)$ for all $g\in G$, and
$\int |\bra \psi |U(g)\vp\ket|^2d\lambda(g) = \|\psi\|^2\|K^{-\frac 12} \vp\|^2$ for all $\vp,\psi\in \hil$, with the understanding that
$\|K^{-\frac 12} \vp\|=\infty$ whenever $\vp\notin D(K^{-\frac 12})$ \cite[Theorem 3]{Duflo}.
However, this holds also in the case of square integrable projective representations (see e.g. \cite[Remark 2]{Cassinelli2}),
which is seen by using the well-known fact that the map $(t,g)\mapsto t^{-1}U(g)$ is an ordinary representation of the group $\T \times_m G$,
where $m$ is the multiplier of $U$.
(Recall that the set $\T \times G$, equipped with the composition $(t,g)(t',g')=(m(g,g')tt',gg')$, becomes a locally compact second
countable topological group, denoted by $\T\times_m G$, whose Borel structure coincides with the product Borel structure
on $\T\times G$; see e.g. \cite[p. 253]{Varadarajan}).
In this case, the formal degree of the projective unitary representation $U$ is defined to be the formal degree of the unitary representation $(t,g)\mapsto t^{-1}U(g)$,
which is clearly square integrable if and only if $U$ is such.

If $U:G\to \hU(\hil)$ is an irreducible square integrable projective unitary representation, we let $C_U$ denote the square root of the formal degree of $U$.
The following Lemma lists the properties of this operator,
in the form which is convenient for our purposes. It is a direct consequence of \cite[Theorem 3]{Duflo}.

\begin{lemma}\label{trlemma} Let $U:G\to \hU(\hil)$ be an irreducible square integrable projective unitary representation.
\begin{itemize}
\item[(a)] Let $S$ be a positive trace class operator, and $\vp\in \hil$. Then
\bet
\int \bra\vp|\beta_U(g)(S)\vp\ket d\lambdat(g) = \int \tr[S\beta_U(g)(|\vp\ket\bra\vp|)]d\lambda(g) = \tr[S]\|C_U^{-1}\vp\|^2,
\eeqt
with the understanding that $\|C^{-1}\vp\|^2=\infty$ whenever $\vp\notin D(C_U^{-1})$.
\item[(b)] $C_U$ and $C_U^{-1}$ are densely defined, selfadjoint, positive, and injective, and satisfy
\beat
U(g)C_U &=& \Delta(g)^{-\frac 12} C_U U(g) \ \ g\in G,\\
U(g)C_U^{-1} &=& \Delta(g)^{\frac 12} C_U^{-1} U(g) \ \ g\in G.
\eeqat
In particular, $U$ leaves $D(C_U)$ and $D(C_U^{-1})$ invariant.
\item[(c)] $C_U$ is bounded if and only if $G$ is unimodular. In that case, $C_U$ is a multiple of the identity.
\end{itemize}
\end{lemma}
\begin{proof}
Assume first that $S=|\psi\ket\bra\psi|$ for some unit vector $\psi\in \hil$. Now
\beat
\int \bra\vp|\beta_U(g)(S)\vp\ket d\lambdat(g) &=& \int \tr[\beta_U(g^{-1})(S)|\vp\ket\bra\vp|] d\lambda(g) = \int \tr[S\beta_U(g)(|\vp\ket\bra\vp|)] d\lambda(g)\\
&=& \int |\bra \psi |U(g)\vp\ket|^2 d\lambda(g).
\eeqat
It follows from Theorem 3 of \cite{Duflo} that the last integral is finite if and only if $\vp\in D(C_U^{-1})$, and is equal to $\|C_U^{-1}\vp\|^2$ in that case.
This proves (a) for $S=|\psi\ket\bra\psi|$. Since the integrands above are all positive, the general case follows by writing
$S=\sum_n t_n |\psi_n\ket\bra\psi_n|$ and using the monotone convergence theorem. This proves (a).
Since the formal degree of $U$ is densely defined, selfadjoint, positive and injective, so are $C_U$ and $C_U^{-1}$ \cite[p. 1189]{DunfordII}.
Since the formal degree of $U$ is the formal degree of the unitary representation $(t,g)\mapsto t^{-1}U(g)$, and the modular function of $\T\times_m G$
is $(t,g)\mapsto \Delta(g)$, Theorem 3 of \cite{Duflo} gives $U(g) C_U^2 = \Delta(g)^{-1}C_U^2U(g)$ for each $g\in G$. By using the spectral representation
of the formal degree $C_U^2$, we see that the first
of the equalities in (b) holds. The second is a consequence of the first and the fact that $U(g)^{-1}=m(g,g^{-1})^{-1}U(g^{-1})$, where $m$ is the
multiplier of $U$. For part (c), see the note following Theorem 3 of \cite{Duflo}.
\end{proof}

The following corollary is a generalization of Lemma 2 of \cite{ncqm}, which no longer holds in the non-unimodular case.

\begin{corollary} Let $U$ be an irreducible square integrable projective representation, and let $S\in \hT(\hil)$ and $A\in L(\hil)$ be positive operators. Then
\bet
\int \tr[A\beta_U(g)(S)] d\lambda(g) = \tr[A]\|C_U^{-1}\sqrt{S}\|_{HS}^2.
\eeqt
(Here we have denoted $\|C_U^{-1}\sqrt{T}\|_{HS}=\infty$ whenever
$C_U^{-1}\sqrt{T}\notin \hHS(\hil)$, and used the conventions $0\cdot \infty = 0$ and $\infty\cdot\infty = \infty$.)
In particular, if $A\neq O$ and $S\neq O$, the function $g\mapsto \tr[A\beta_U(g)(S)]$ is $\lambda$-integrable if and only if
$A\in \hT(\hil)$ and $C_U^{-1}\sqrt{S}\in \hHS(\hil)$. 
\end{corollary}
\begin{proof}
Clearly the case where either $A=O$ or $S=O$ is trivial, so we can proceed by assuming that both are nonzero.
\begin{enumerate}
\item If $A\in \hT(\hil)$ is positive and $S=|\eta\ket\bra\eta|$, with $\eta\in \hil$ a unit vector, we are in the situation of Lemma \ref{trlemma} (a).
\item Let $A,S\in\hT(\hil)$ be positive and nonzero. 
Let $(w_n)$ be a sequence of nonnegative numbers and $(\eta_n)$ an orthonormal sequence of vectors such that
$S = \sum_n w_n |\eta_n\ket\bra\eta_n|$. Since each $\beta_U(g)$ is trace norm continuous, it follows from Lemma \ref{trlemma} (a) and
the monotone convergence theorem that $\int \tr[A\beta_U(g)(S)] d\lambda(g) = \tr[A]M$, where $M=\sum_n w_n \|C_U^{-1}\eta_n\|^2$
(with the understanding that $\|C_U^{-1}\eta_n\|=\infty$ whenever $\eta_n\notin D(C_U^{-1})$).

If $M=\infty$, then $C_U^{-1}\sqrt{S}$ cannot be a Hilbert-Schmidt operator, since
otherwise $\|C_U^{-1}\sqrt{S}\|_{HS}^2 = \sum_{\xi\in\hK} \|C^{-1}\sqrt{S}\xi\|^2 = M<\infty$, where $\hK$ is an orthonormal basis including
all the $\eta_n$. Hence $\|C_U^{-1}\sqrt{S}\|_{HS}^2=\infty = M$.
 
Assume then that $M<\infty$, so that, in particular, $\eta_n\in D(C_U^{-1})$ for all those $n\in\N$ for which $w_n>0$. Let $\vp\in\hil$. Since
the series $\sqrt{S} = \sum_n \sqrt{w_n} |\eta_n\ket\bra\eta_n|$ converges in the operator norm,
the vector series $\sum_n \sqrt{w_n} \bra \eta_n |\vp\ket \eta_n$ converges to $\sqrt{S}\vp$ in the norm of
$\hil$. Since $(\eta_n)$ is orthonormal, the Cauchy-Schwarz inequality gives
\bet
\sum_n \sqrt{w_n} |\bra \eta_n |\vp\ket| \|C_U^{-1}\eta_n\|\leq \sqrt{M} \|\vp\|<\infty,
\eeqt
so also the series $\sum_n \sqrt{w_n} \bra \eta_n |\vp\ket C_U^{-1}\eta_n$ converges in $\hil$. Since $C_U^{-1}$ is closed by Lemma \ref{trlemma} (b), it follows
that $\sqrt{S}\vp\in D(C_U^{-1})$ and $C_U^{-1}\sqrt{S}\vp$ equals the sum of the latter series. In particular, $D(C_U^{-1}\sqrt{S}) =\hil$.
Now the previous inequality shows that $\|C_U^{-1}\sqrt{S}\vp\|\leq \sqrt{M} \|\vp\|$, so $C_U^{-1}\sqrt{S}$ is bounded.
Clearly $\|C_U^{-1}\sqrt{S}\|_{HS}^2= \sum_{\xi\in\hK} \|C_U^{-1}\sqrt{S}\xi\|^2 = M<\infty$ if $\hK$ is an orthonormal basis of $\hil$ which includes all
the $\eta_n$, so $C_U^{-1}\sqrt{S}$ is a Hilbert-Schmidt operator, with $\|C_U^{-1}\sqrt{S}\|_{HS}^2=M$.

\item We are left with the general case, with $A$ and $S$ nonzero.
By repeating the steps 3 and 4 in the proof of Lemma 2 of \cite{ncqm} (with obvious alterations), we see that the Corollary is true in this case also.
(Notice that this includes the possibility that $C_U^{-1}\sqrt{S}$ is not a Hilbert-Schmidt operator.)
The proof is complete.
\end{enumerate}
\end{proof}

\section{Covariant observables and normal covariant maps}\label{ncm}

Now we are ready to proceed to the actual setting.
We fix $U:G\to \hU(\hil)$ to be an irreducible projective unitary representation, and let $\beta: G\to \aut(\hT(\hil))$ denote the
associated map $\beta_U$. (According to the Remark following that lemma, in the case
where $G$ is connected, it would be equivalent to start with a continuous group homomorphism $\beta$, with the property \eqref{irrcond},
as was done in \cite{ncqm}.) The projective representation $U$ and the associated map $\beta$ will remain fixed throughout the rest of the paper.

The following definition for covariance was used in \cite{ncqm}.

\

\noindent {\bf Definition.}
\begin{itemize}
\item[(a)] A linear map $\Gamma:L^{\infty}(G, \lambda)\to L(\hil)$ is \emph{$\beta$-covariant}, if
$\beta(g)^*(\Gamma(f)) =\Gamma(f(g\cdot))$ for all $f\in L^\infty(G, \lambda)$, $g\in G$. 
\item[(b)] An observable $E:\hB(G)\to L(\hil)$ is \emph{$\beta$-covariant} if
$\beta(g)^*(E(B)) = E(g^{-1}B)$ for all $g\in G$ and $B\in \hB(G)$.
\end{itemize}

\

As mentioned in \cite{ncqm}, each \emph{normal} (i.e. weak-* continuous) linear positive $\beta$-covariant map
$\Gamma:L^{\infty}(G, \lambda)\to L(\hil)$, with $\Gamma(g\mapsto 1)=I$, defines a $\beta$-covariant observable
$B\mapsto \Gamma(\chi_B)$. Conversely, each $\beta$-covariant observable $E$ gives rise to a normal linear map
$\Gamma:L^{\infty}(G, \lambda)\to L(\hil)$ via the ultraweak operator integrals $\Gamma(f) = \int f dE$, where $\vp,\psi\in \hil$.
The proof of the latter statement was given in \cite{ncqm} in the unimodular case, but it holds also in general. (See the discussion preceding
Lemma 6 of \cite{ncqm}, and notice that part (a) of the lemma does not use unimodularity.)

Thus, normal covariant maps are essentially the same as covariant observables. In \cite{ncqm}, normal covariant maps were used
in proving the characterization. Here we use observables.

Our main result, Theorem 2, has the following, easily proved converse.
\begin{theorem}\label{obstheorem} Assume that $U$ is square integrable, and
let $S$ be a positive operator of trace one. Then there is a $\beta$-covariant observable $E:\hB(G)\to L(\hil)$, such that
\bet
\bra \vp |E(B)\psi\ket = \int_B \bra C_U\vp |\beta(g) (S) C_U\psi\ket d\lambdat(g), \ \ \ B\in\hB(G), \vp,\psi\in D(C_U).
\eeqt
\end{theorem}
\begin{proof}
It follows from Lemma \ref{trlemma} (a) and the polarization identity that the integrals of the above form exist. Let $B\in \hB(G)$, and
define a symmetric sesquilinear form $\Phi_B:D(C_U)\times D(C_U)\to \C$ by
\bet
\Phi_B(\psi, \vp) = \int_B \bra C_U\psi |\beta(g) (S) C_U\vp\ket d\lambdat(g).
\eeqt
It follows from Lemma \ref{trlemma} (a) that $0 \leq \Phi_B(\vp,\vp)\leq \|\vp\|^2$ for all $\vp\in D(C_U)$.
Hence, by using the polarization identity and the density of
$D(C_U)$, we can extend $\Phi_B$ to a bounded symmetric sesquilinear form defined on $\hil\times\hil$. Thus, there is a selfadjoint operator
$E(B)\in L(\hil)$, such that $\bra \psi |E(B)\vp\ket = \int_B \bra C_U\vp |\beta(g) (S) C_U\psi\ket d\lambdat(g)$ for all $\psi, \vp\in D(C_U)$.
Moreover, we have $0\leq E(B)\leq I$, $E(\emptyset)=O$ and $E(G)=I$.
It follows from the $\sigma$-additivity of the indefinite integral that
$B\mapsto \bra \vp |E(B)\vp\ket$ is a positive measure for each $\vp\in D(C_U)$. Since $D(C_U)$ is dense and $\|E(B)\|\leq 1$ for each $B\in\hB(G)$,
it follows that $B\mapsto \bra \vp |E(B)\vp\ket$ is a positive measure for each $\vp\in \hil$.

Indeed, let $\vp\in\hil$,
and $(\vp_n)$ be a sequence of vectors in $D(C_U)$ converging to $\vp$. It is clear that $B\mapsto \bra \vp |E(B)\vp\ket$ becomes additive.
Let $(B_k)$ a decreasing sequence of sets in $\hB(G)$ with empty intersection.  Since $\|E(B_k)\|\leq 1$ for all $k$,
the limit $\lim_n \bra \vp_n |E(B_k)\vp_n\ket = \bra\vp |E(B_k)\vp\ket$ exists uniformly for $k\in\N$, so that
\bet
\lim_k \bra \vp|E(B_k)\vp\ket = \lim_k \lim_n \bra \vp_n|E(B_k)\vp_n\ket = \lim_n \lim_k \bra \vp_n|E(B_k)\vp_n\ket = 0,
\eeqt
implying that $\bra\vp|E(\cdot)\vp\ket$ is a positive measure.

Hence, $B\mapsto E(B)$ is a positive normalized operator measure. We are left to prove that $E$ is $\beta$-covariant.
Take $h\in G$, $B\in \hB(G)$, and $\vp\in D(C_U)$. By using Lemma \ref{trlemma} (b) and the left invariance of $\lambda$, we get
\beat
\bra \vp|\beta(h^{-1})^*(E(B))\vp\ket &=& \bra U(h^{-1})\vp|E(B)U(h^{-1})\vp\ket = \int_B \bra C_UU(h^{-1})\vp|\beta(g)(S)C_UU(h^{-1})\vp\ket d\lambdat(g)\\
&=& \int_B \Delta(h^{-1})\bra U(h^{-1})C_U\vp|\beta(g)(S)U(h^{-1})C_U\vp\ket d\lambdat(g) \\
&=& \int_B \Delta(h^{-1})\bra C_U\vp|\beta(h^{-1})^{-1}(\beta(g)(S))C_U\vp\ket d\lambdat(g)\\
&=& \int_B \bra C_U\vp|\beta(hg)(S)C_U\vp\ket \Delta((hg)^{-1})d\lambda(g)\\
&=& \int_{hB} \bra C_U\vp|\beta(g)(S)C_U\vp\ket \Delta(g^{-1})d\lambda(g)\\
&=& \int_{hB} \bra C_U\vp|\beta(g)(S)C_U\vp\ket d\lambdat(g) = \bra \vp |E(hB)\vp\ket,
\eeqat
proving the covariance of $E$.
\end{proof}

Theorem \ref{obstheorem} states that when the projective representation $U$ is square integrable, there exist $\beta$-covariant observables.
Part (b) of the following proposition gives the interesting fact that the converse is also true: the existence of a $\beta$-covariant observable
implies the square integrability of $U$. This result is contained in \cite[Theorem 2]{Cassinelli}, proved using the generalized imprimitivity theorem.
Here we give a simple direct proof which uses only the properties of Haar measures (and basic operator theory).

\begin{proposition}\label{squareintlemma} Assume that there is a $\beta$-covariant observable $E:\hB(G)\to L(\hil)$.
\begin{itemize}
\item[(a)] If $B\in \hB(G)$, then
\bet
\lambdat(B) = \int \tr[\beta(g^{-1})^*(E(B)) S] d\lambda(g) 
\eeqt
for all positive operators $S$ of trace one.
\item[(b)] The irreducible projective representation $U$ is square-integrable.
\end{itemize}
\end{proposition}
\begin{proof}
Let $B\in \hB(G)$, let $S\in \hT(\hil)$ be positive and of trace one, and let $\mu$ denote
the probability measure $B\mapsto \tr[E(B)S]$. Now
\beat
\lambdat(B) &=& \lambda(B^{-1}) = \int\left(\int \chi_{B^{-1}}(g)d\lambda(g)\right) d\mu(g')=\int\left(\int \chi_{B^{-1}}((g')^{-1}g)d\lambda(g)\right) d\mu(g') \\
&=& \int\left(\int \chi_{B^{-1}}((g')^{-1}g)d\mu(g')\right) d\lambda(g) = \int \left(\int_{gB} d\mu(g')\right)d\lambda(g) \\
&=& \int \tr[E(gB)S] d\lambda(g) = \int \tr[\beta(g^{-1})^*(E(B))S] d\lambda(g),
\eeqat
where the left invariance of $\lambda$, Fubini's theorem and the $\beta$-covariance of $E$ have been used. This proves (a).
To prove (b), take any $B\in \hB(G)$, such that $0<\lambdat(B)<\infty$. Since (a) holds, we have $E(B)\neq 0$, so by the spectral theorem, there is a nonzero projection
$P$, and a real number $t>0$, such that $tP\leq E(B)$. Take any unit vector $\vp\in P(\hil)$. Now $t|\vp\ket\bra \vp|\leq E(B)$, so by using (a) we get
\bet
t \int |\bra \vp|U(g)\vp\ket|^2 d\lambda(g) = t \int \tr[|\vp\ket\bra \vp|\beta(g^{-1})(|\vp\ket\bra\vp|)] d\lambda(g) \leq \lambdat(B) <\infty,
\eeqt
proving (b).
\end{proof}

From now on, we assume that there exists a $\beta$-covariant observable, so $U$ is square integrable by Proposition \ref{squareintlemma} (b).
For simplicity, we let $C$ denote the operator $C_U$.

\begin{lemma}\label{hslemma} Let $E:\hB(G)\to L(\hil)$ be a $\beta$-covariant observable. Then
\bet
\lambdat(B) = \|C^{-1}E(B)^{\frac 12}\|_{HS}^2, \ \ B\in \hB(G),
\eeqt
where it is understood that $\|C^{-1}E(B)^{\frac 12}\|_{HS}=\infty$ whenever
$C^{-1}E(B)^{\frac 12}\notin \hHS(\hil)$.
\end{lemma}
\begin{proof}
Let $\vp\in \hil$ be a unit vector, and $\{\vp_n\}$ an orthonormal basis of $\hil$ containing $\vp$. Let $\psi\in \hil$ be any unit vector.
Now Proposition \ref{squareintlemma} (a), the monotone convergence theorem, and Lemma \ref{trlemma} (a) give
\beat
\lambdat(B) &=& \int \tr[\beta(g^{-1})^*(E(B))|\psi\ket\bra\psi|] d\lambda(g) = \int \tr[E(B)\beta(g^{-1})(|\psi\ket\bra\psi|)] d\lambda(g) \\
&=& \int \tr[E(B)^{\frac 12}\beta(g)(|\psi\ket\bra\psi|)E(B)^{\frac 12}] d\lambdat(g) \\
&=& \sum_{n=1}^\infty \int \bra E(B)^{\frac 12}\vp_n |\beta(g)(|\psi\ket\bra\psi|)E(B)^{\frac 12}\vp_n\ket d\lambdat(g) \\
&=& \sum_{n=1}^\infty \|C^{-1}E(B)^{\frac 12}\vp_n\|^2.
\eeqat
This holds regardless of whether $\lambdat(B)$ is finite or not, with the understanding that $\|C^{-1}E(B)^{\frac 12}\vp_n\|=\infty$
if $E(B)^{\frac 12}\vp_n\notin D(C^{-1})$.

If $\lambdat(B)=\infty$, the above calculation shows that $C^{-1}E(B)^{\frac 12}\notin \hHS(\hil)$ (it need not even
be bounded). Thus then $\lambdat(B) = \infty = \|C^{-1}E(B)^{\frac 12}\|_{HS}^2$.

Assume now that $\lambdat(B)<\infty$. It follows that the above series converges, so, in particular, $E(B)^{\frac 12}\vp_n\in D(C^{-1})$
for all $n\in\N$. Since the basis was chosen to contain the arbitrarily picked unit vector $\vp\in\hil$, we get
$D(C^{-1}E(B)^{\frac 12})=\hil$. Since $C^{-1}$ is selfadjoint, it is closed, so also $C^{-1}E(B)^{\frac 12}$ is closed (because $E(B)^{\frac 12}$ is bounded).
Since the domain of $C^{-1}E(B)^{\frac 12}$ is all of $\hil$, it follows by the closed graph theorem that $C^{-1}E(B)^{\frac 12}$ is bounded.
The above calculation now shows that it is of the Hilbert-Schmidt class, with $\lambdat(B) = \|C^{-1}E(B)^{\frac 12}\|_{HS}^2$.
The proof is complete.
\end{proof}
\noindent {\bf Remark. } Notice that this result generalizes the relation $\tr[E(B)] = d^{-1}\lambda(B)$, which
holds in the unimodular case, with $d^{-1}I$ the formal degree (see the proof of Lemma 6 (b) of \cite{ncqm}).

\

The next Theorem states that every $\beta$-covariant observable is of the form of Theorem \ref{obstheorem} for some
positive operator $S$ of trace one. This is analogous to Theorem 3 of \cite{ncqm}, where the formal degree is $d^{-1} I$.
By using Lemma \ref{trlemma} (b) and the fact that
''$d\lambda(g) = \Delta(g)d\lambdat(g)$'', we see that this characterization is indeed the same as that of \cite[Theorem 2]{Cassinelli}.
While the proof of \cite[Theorem 2]{Cassinelli} is based on the generalized imprimitivity theorem, the following proof relies on the
fact that as a separable dual space, $\hT(\hil)$ has the Radon-Nikod\'ym property \cite[p. 79]{Diestel}.

\begin{theorem}\label{observables}
Let $E:\hB(G)\to L(\hil)$ be a $\beta$-covariant observable. Then there is a unique
positive operator $S\in \hT(\hil)$ of trace one, such that
\be\label{repfinal}
\bra \vp |E(B)\psi\ket = \int_B \bra C\vp |\hog{g} (S) C\psi\ket d\lambdat(g), \ \ \ B\in\hB(G), \vp,\psi\in D(C).
\eeq
\end{theorem}
\begin{proof}
Let $B\in\hB(G)$ be such that $\lambdat(B)<\infty$, and let $E_B$ be the Hilbert-Schmidt operator $C^{-1}E(B)^{\frac 12}$ (see
Lemma \ref{hslemma}). Now define $A(B)= E_BE_B^*$, so that $A(B)$ is a positive trace class operator. If
$\{\psi_n\}$ is any orhonormal basis of $\hil$, we have
\bet
\tr[A(B)] = \sum_n \bra \psi_n |A(B)\psi_n\ket = \sum_n \|E_B^*\psi_n\|^2 = \|E_B^*\|_{HS}^2 = \|E_B\|_{HS}^2 = \lambdat(B)
\eeqt
by Lemma \ref{hslemma}. Thus
\be\label{continuity}
\tr[A(B)] = \lambdat(B), \ \ \ B\in \hB(G), \lambdat(B)<\infty.
\eeq

Now $E(B)^{\frac 12}C^{-1}\subset E_B^*$ because $C^{-1}$ is selfadjoint, so that
\be\label{AE}
A(B)\vp = E_BE_B^*\vp=C^{-1}E(B)C^{-1}\vp \ \ \text{ for each } \vp\in D(C^{-1}).
\eeq
Let $B\in \hB(G)$, $\lambdat(B)<\infty$.
Let $h\in G$. Now also $\lambdat(hB)=\Delta(h^{-1})\lambdat(B)<\infty$. Let $\vp\in D(C^{-1})$.
By Lemma \ref{trlemma} (b), $U(h)^*\vp=m(h,h^{-1})U(h^{-1})\vp\in D(C^{-1})$, where $m$ is the multiplier of $U$.
Using covariance and Lemma \ref{trlemma} (b), we get
\beat
A(hB)\vp &=& C^{-1}E(hB)C^{-1}\vp = C^{-1}U(h)E(B)U(h)^*C^{-1} = \Delta(h)^{-1}U(h)C^{-1}E(B)C^{-1}U(h)^*\vp\\
&=& \Delta(h)^{-1}U(h)A(B)U(h)^*\vp = \Delta(h)^{-1}\beta(h)(A(B))\vp.
\eeqat
Since $A(hB)$ and $\beta(h)(A(B))$ are bounded, and $D(C^{-1})$ is dense, we get
\be\label{covariance}
A(hB) = \Delta(h)^{-1}\beta(h)(A(B)), \ \ h\in G, \ B\in \hB(G), \lambdat(B)<\infty.
\eeq

Now we can proceed in much the same way as in the proof of Proposition 1 of \cite{ncqm}. For any $D\in \hB(G)$, we let
$\hB(D)$ denote the $\sigma$-algebra $\{ B\cap D \mid B\in \hB(G)\}$. If $D\in \hB(G)$ is such that $\lambdat(D)<\infty$, then
\eqref{continuity} implies that $A(B)\in \hT(\hil)$ for each $B\in \hB(D)$, so we have a set function $\mu_D:\hB(D)\to \hT(\hil)$,
defined by $B\mapsto A(B)$. This set function is additive, with $\mu_D(\emptyset)=O$, since $E$ is an operator measure and \eqref{AE}
holds. Moreover, \eqref{continuity} implies that $\mu_D$ is $\sigma$-additive with respect to the trace norm, i.e. $\mu_D$ is
a $\hT(\hil)$-valued vector measure.

Since $\lambdat$ is $\sigma$-finite, we can write $G=\bigcup_{n\in\N} K_n$ with $(K_n)$ a sequence of disjoint
sets in $\hB(G)$ of finite $\lambdat$-measure. 
Denote by $\lambdat_n$ the restriction of $\lambdat$ to the $\sigma$-algebra $\hB(K_n)$, and let $\mu_n = \mu_{K_n}$ for each $n\in \N$.
Now \eqref{continuity} implies that the vector measure $\mu_n$ is a $\lambdat_n$-continuous, with the variation $|\mu_n|$ bounded and given by
\bet
|\mu_n|(B) := \sup \left\{ \sum_{D\in \pi} \|A(D)\|_{\tr} \mid \pi \text{ a finite disjoint partition of } B\right\} = \lambdat(B)
\eeqt
for all $B\in \hB(K_n)$. (See \cite[pp. 1-2, 11]{Diestel} for the definitions).

Since each $\mu_n$ is $\lambdat_n$-continuous and of bounded variation,
and each measure $\lambdat_n$ is finite, it follows from the Radon-Nikodym property of $\hT(\hil)$ \cite[p. 79]{Diestel} that for each $n\in\N$ there is a
$\lambdat_n$-integrable function $v_n:K_n\to \hT(\hil)$, such that $A(B) = \mu_n(B) =\int_B v_n d\lambdat_n$ for all $B\in \hB (K_n)$.
Moreover, $\lambdat(B) = |\mu_n|(B) = \int_B \|v_n(g)\|_{\tr} \ d\lambdat_n(g)$ \cite[p. 46]{Diestel}, so that
$\|v_n(g)\|_{\tr} = 1$ for $\lambdat$-almost all $g\in K_n$ \cite[Corollary 5, p. 47]{Diestel}. Since $A(B)$ is positive for all
$B\in \hB(K_n)$, we have that for each $\vp\in \hil$, there is a null set $N_{\vp}\subset K_n$, such that $\bra \vp |v_n(g)\vp\ket\geq 0$ for all $g\in K_n\setminus N_{\vp}$.
Since $\hil$, being separable, contains a countable dense subset, it follows that for almost all $g\in K_n$, $\bra \vp|v_n(g)\vp\ket\geq 0$ for all $\vp\in \hil$,
which means that $v_n(g)$ is a positive operator for almost all $g\in K_n$. Thus $v_n(g)$ is a positive operator of trace one for almost all $g\in K_n$.

The function $v_n$ in the representation $\mu_n(B) =\int_B v_n d\lambdat_n$ is $\lambdat_n$-essentially unique
by \cite[Corollary 5, p. 47]{Diestel}.

For each $n\in\N$, we denote by $v_n$ also the function defined on the whole of $G$ which coincides with $v_n$ in $K_n$
and vanishes elsewhere. Let $v=\sum_{n=1}^\infty v_n$ (pointwise). Since $v$ is a pointwise limit of $\lambdat$-measurable functions, it
is itself $\lambdat$-measurable. (Note that since the $K_n$ are disjoint, $v(g) = v_n(g)$ for $g\in K_n$.) Clearly
\be\label{norm}
\|v(g)\|_{\tr} = 1 \ , v(g) \geq 0 \ \ \text{ for almost all } g\in G.
\eeq
Consequently, (see \cite[Theorem 2, p. 45]{Diestel}), $v$ is $\lambdat$-integrable over any set $B\in\hB(G)$ of finite $\lambdat$-measure.
Now let $B\in \hB(G)$ be such that $\lambdat(B)<\infty$. For any $k\in\N$, we have
\bet
\int_{\cup_{n=1}^k K_n} \chi_B v d\lambdat = \sum_{n=1}^k \int_{B\cap K_n} v_n d\lambdat_n = \sum_{n=1}^k A(B\cap K_n) = \sum_{n=1}^k \mu_B(B\cap K_n),
\eeqt
so by the $\sigma$-additivity of the indefinite integral of $\chi_Bv$,
the series $\sum_{n=1}^\infty \mu_B(K_n)$ (of trace class operators) converges in the trace norm to (the trace class operator)
$\int_B v d\lambdat$. On the other hand, also $\mu_B$ is $\sigma$-additive with respect to the trace norm, so this series converges to
$\mu_B(B) = A(B)$. Hence,
\be\label{rep}
A(B) = \int_B v d\lambdat \ \ \text{ for all } B\in \hB(G) \text{ with } \lambdat(B)<\infty.
\eeq
The function $v$ in this representation is clearly $\lambdat$-essentially uniquely determined.

Next, let $B\in \hB(G)$ be such that $\lambdat(B)<\infty$, and $h\in G$. Now also $\lambdat(hB)=\Delta(h^{-1})\lambdat(B)<\infty$.
By the left invariance of $\lambda$, \eqref{rep}, \eqref{covariance}, and the fact that $\beta(h)$ is a trace norm continuous linear map, we get
\beat
\Delta(h^{-1})\int_B v(hg)d\lambdat(g) &=& \int_B v(hg)\Delta(g^{-1})\Delta(h^{-1})d\lambda(g) = \int \chi_B(g) v(hg)\Delta((hg)^{-1})d\lambda(g)\\
&=& \int \chi_B(h^{-1}g)v(g)\Delta(g^{-1})d\lambda(g) = \int_{hB} v d\lambdat = A(hB)\\
&=& \Delta(h^{-1})\beta(h)(A(B)) = \Delta(h^{-1})\int_B \beta(h)(v(g))d\lambdat(g).
\eeqat
Since $\lambdat$ is $\sigma$-finite, this implies (by \cite[Corollary 5, p. 47]{Diestel}) that
\be\label{vcov}
\text{ for each } h\in G, \ \ \beta(h)(v(g)) = v(hg) \ \ \text{ for }\lambda\text{-almost all } g\in G.
\eeq
Here we have used also the fact that $\lambdat$ and $\lambda$ have the same null sets.
Define the function $v_0:G\to \hT(\hil)$ by $v_0(g) = \beta(g^{-1}) (v(g))$. Then $v_0$ is $\lambda$-measurable by Lemma
\ref{measurabilitylemma}.
Let $h\in G$. Now $v(g) = \beta(g)(v_0(g))$ for each $g$, so using \eqref{vcov}, we get
\bet
\beta(h)(\beta(g)(v_0(g))) = \beta(h)(v(g)) = v(hg) = \beta(hg)(v_0(hg)),
\eeqt
for $\lambda$-almost all $g$, from which it follows (since $\beta$ is a homomorphism) that
\be\label{vcov2}
\text{ for each } h\in G, \ \ v_0(g) = v_0(hg) \ \ \text{ for }\lambda\text{-almost all } g\in G.
\eeq
In addition, since each $\beta(g^{-1})$ preserves the trace norm, it follows from \eqref{norm} that
$v_0$ is a $\lambdat$-essentially bounded function, so Lemma 4 of \cite{ncqm} can be applied to get an $S\in \hT(\hil)$, such that
$v_0(g) = S$ for $\lambda$-almost all $g\in G$, i.e. $v(g) = \beta(g)(S)$ for $\lambda$-almost all $g\in G$. It now follows from \eqref{rep} that
\be\label{rep2}
A(B) = \int_B \beta(g)(S) d\lambdat(g) \ \ \text{ for all } B\in \hB(G) \text{ with } \lambdat(B)<\infty.
\eeq
Since the function $v$ was $\lambdat$-essentially unique in the representation \eqref{rep}, it follows by
\cite[Corollary 5, p. 47]{Diestel} that $S$ in the representation \eqref{rep2} is uniquely determined. Since $v(g)$ is positive and of trace one
for almost all $g\in G$, and each $\beta(g^{-1})$ preserves positivity and the trace norm, we see that $S$ must be a positive operator of trace one.

Next, let $B\in \hB(G)$ be arbitrary, and let $\vp\in D(C^{-1})$. Denote $T=|\vp\ket\bra\vp|$. We have 
\beat
\bra C^{-1}\vp |E(B)C^{-1}\psi\ket &=& \sum_{n=1}^\infty \tr[TA(B\cap K_n)]
= \sum_{n=1}^\infty \int_{B\cap K_n} \tr[T\beta(g)(S)] d\lambdat(g)\\
&=& \int_B\tr[T\beta(g)(S)] d\lambdat(g),
\eeqat
where the first equality follows since $\bra C^{-1}\vp |E(\cdot)C^{-1}\vp\ket$ is a measure, the second is given by \eqref{rep2} and the fact that
$\hT(\hil)\ni V\mapsto \tr[TV]\in \C$ is trace norm continuous, and the last is due to the
$\sigma$-additivity of the indefinite integral. Note that to get the last equality, we need the fact that the function
$g\mapsto \tr[T\beta(g)(S)]$ is $\lambdat$-integrable by Lemma \ref{trlemma} (a). The relation \eqref{repfinal}
now follows by polarization.

We are left to show that $S$ is unique in the representation \eqref{repfinal}. 
Assume that also $S'\in\hT(\hil)$ is positive and of trace one,
and satisfies \eqref{repfinal}. Let $B\in \hB(G)$ be of finite $\lambdat$-measure. Since $\|\beta(g)(S')\|_{\tr}=\|S'\|_{\tr}= 1$ for all $g$, the function
$G\ni g\mapsto \chi_B\beta(g)(S')\in \hT(\hil)$ (which is $\lambda$-, and hence $\lambdat$-measurable by Lemma \ref{measurabilitylemma}) is
$\lambdat$-integrable by \cite[Theorem 2, p. 45]{Diestel}. Hence, for each $\vp\in D(C^{-1})$, we get,
\bet
\bra\vp|A(B)\vp\ket = \bra C^{-1} \vp |E(B) C^{-1}\vp\ket = \int_B \bra\vp |\beta(g)(S')\vp\ket d\lambdat(g) = \tr\left[|\vp\ket\bra\vp| \left(\int_B \beta(g)(S')d\lambdat(g)\right)\right]
\eeqt
(by using also the fact that the functional $\hT(\hil)\ni V\mapsto \tr[|\vp\ket\bra\vp|V]\in \C$ is trace norm continuous), so
$A(B) = \int_B \beta(g)(S')d\lambdat(g)$. Now by the uniqueness of $S$ in the representation \eqref{rep2} it follows that $S=S'$.
The proof is complete.
\end{proof}
\noindent {\bf Remark. } 
According to the discussion in the beginning of the present Section, Theorem \ref{observables} gives the following characterization
for normal covariant maps:
Let $\Gamma:L^{\infty}(G, \lambda)\to L(\hil)$ be linear, positive, normal, and $\beta$-covariant, with $\Gamma(g\mapsto 1) = I$. Then
there is a unique positive operator $S$ of trace one, such that
\bet
\bra \vp|\Gamma(f)\psi\ket = \int f(g)  \bra C\vp |\beta(g) (S) C\psi\ket d\lambdat(g), \ \ \ \vp,\psi\in D(C), f\in L^{\infty}(G, \lambda).
\eeqt
This should be compared with Theorem 2 of \cite{ncqm}.

\

\noindent{\bf Acknowledgment.} The author thanks Dr. Gianni Cassinelli, Dr. Pekka Lahti and Dr. Kari Ylinen for fruitful discussions and valuable comments
on the paper.

\


\begin{thebibliography}{99}
\bibitem{Cassinelli2} G. Cassinelli, G. D'Ariano, E. De Vito, A. Levrero, Group theoretical quantum tomography, J. Math. Phys. {\bf 41} 7940-7951 (2000).
\bibitem{Symmetry} G. Cassinelli, E. De Vito, P. J. Lahti, A. Levrero, {\em The Theory of Symmetry Actions in Quantum Mechanics
With an Application to the Galilei Group}, Springer, Berlin, 2004.
\bibitem{Cassinelli} G. Cassinelli, E. De Vito, A. Toigo, Positive operator valued measures covariant with respect to an irreducible representation,
{\em J. Math. Phys.} {\bf 44} 4768-4775 (2003).
\bibitem{Davies} E. B. Davies, {\em Quantum Theory of Open Systems}, Academic Press, London, 1976.
\bibitem{Diestel} J. Diestel, J. J. Uhl, Jr., {\em Vector Measures (Mathematical Surveys 15)}, American Mathematical Society,
Providence, 1977.
\bibitem{Duflo} M. Duflo, C. Moore, On the regular representation of a nonunimodular locally compact group,
{\em J. Funct. Anal.} {\bf 21} 209-243 (1976).
\bibitem{Dunford} N. Dunford, J. T. Schwartz, {\em Linear Operators, Part I: General Theory}, Interscience Publishers,
New York, 1958.
\bibitem{DunfordII} N. Dunford, J. T. Schwartz, {\em Linear Operators, Part II: Spectral Theory}, Interscience Publishers,
New York, 1964.
\bibitem{Holevo} A. S. Holevo, Covariant measurements and uncertainty relations, {\em Rep. Math. Phys. }{\bf 16} 385-400 (1979).
\bibitem{HolevoII} A. S. Holevo, {\em Statistical Structure of Quantum Theory}, Springer, Berlin, 2001.
\bibitem{ncqm} J. Kiukas, P. Lahti, K. Ylinen, Normal covariant quantization maps, {\em J. Math. Anal. Appl.}, in press.
\bibitem{Varadarajan} V. S. Varadarajan, \emph{Geometry of quantum theory}, Second Edition, Springer-Verlag, New York, 1985.
\bibitem{Werner} R. Werner, Quantum harmonic analysis on phase space, {\em J. Math. Phys.} {\bf 25} 1404-1411 (1984).
\end{thebibliography}
\end{document}